\documentclass[a4paper,11pt]{article}
\pdfoutput=1
\usepackage{jcappub}
\usepackage{graphicx}
\usepackage{color}
\usepackage{amssymb}
\usepackage{amsmath}
\usepackage{mathtools}
\usepackage{url}
\usepackage{natbib}
\usepackage[normalem]{ulem}
\usepackage{afterpage}
\usepackage{float}
\usepackage{tensor}

\newcommand{\bn}{{\mathbf n}}
\newcommand{\bv}{{\mathbf v}}

\newcommand{\bnabla}{{\boldsymbol \nabla}}

\newcommand{\LL}{{\cal L}}

\newcommand{\HH}{{\cal H}}

\newcommand{\SSS}{{\cal S}}

\newcommand{\de}{\delta}
\newcommand{\De}{\Delta}

\newcommand{\ga}{\gamma}

\newcommand{\La}{\Lambda}

\newcommand{\be}{\begin{equation}}
\newcommand{\ee}{\end{equation}}

\newcommand{\bea}{\begin{eqnarray}}
\newcommand{\eea}{\end{eqnarray}}
\newcommand{\bean}{\begin{eqnarray*}}
\newcommand{\eean}{\end{eqnarray*}}

\newcommand{\etac}{t}

\newcommand\spart{\;\raise1.0pt\hbox{$/$}\hskip-6pt\partial}
\definecolor{dred}{rgb}{0.8,0,0.1}
\definecolor{orange}{rgb}{1,0.5,0}

\title{Cosmological Number Counts in Einstein and Jordan frames}
\author[]{J\'er\'emie Francfort,}
\author[]{Basundhara Ghosh and}
\author[]{Ruth Durrer}
\affiliation[]{D\'epartement de Physique Th\'eorique and Center for Astroparticle Physics,\\
Universit\'e de Gen\`eve, 24 quai Ernest  Ansermet, 1211 Gen\`eve 4, Switzerland}

\emailAdd{jeremie.francfort@unige.ch}
\emailAdd{basundhara.ghosh@unige.ch}
\emailAdd{ruth.durrer@unige.ch}

\abstract{Even though we know that physical observations are frame independent, the frame dependence of cosmological perturbations is relatively subtle and has led to confusion in the past. In this paper we show that while the (unobservable) matter power spectrum is frame dependent, the observable number counts are not. We shall also determine how the frame dependence of the power spectrum depends on scale.}


\arxivnumber{}

\begin{document}
\maketitle
\flushbottom

\section{Introduction}\label{s:intro}
From the point of view of fundamental physics the problem of dark energy is very puzzling. Cosmological data, especially the accelerated expansion of the Universe, can be fit relatively well with the standard $\La$CDM cosmological model. However, the value of the cosmological constant $\La$ corresponds to a vacuum energy density
\be
\rho_\La = \frac{\La}{8\pi G} \simeq \left(3\times 10^{-3}{\rm eV}\right)^4
\ee
which cannot be explained by any fundamental theory. Assuming that the vacuum energy scale should be of the order of the cutoff scale of the theory and setting this scale to the Planck scale, one finds that the cosmological constant proposed in the $\La$CDM model is about 120 orders of magnitude smaller than this naively guessed value. Probably the worst guess in the history of physics.

Of course the  cosmological constant cannot be computed in quantum field theory
and it acquires very large corrections at each order in perturbation theory, but the fact that the measured value is so much smaller than any naive guess, is suggestive of the fact that there might be a theoretical reason (maybe coming from quantum gravity) which requests it to vanish. In this case, the observed cosmic acceleration which has led to its introduction must have another origin.\par
This and similar ideas have prompted many workers in the field to consider theories of gravity which modify Einstein gravity in the infrared regime.  The simplest modifications are the so-called scalar-tensor theories which allow, in addition to the metric, for a scalar field with a universal coupling to matter.  These theories can be formulated equivalently in the so-called Einstein or Jordan frame.\par
It is clear that theories which can be transformed into each other by a pure conformal frame transformation describe the same physics and therefore the predicted outcome for every experiment must be equal. However, the interpretation 
of the experiment might be very different: for example, starting from `ordinary' cosmology described by Einstein's theory of gravity leading to an expanding universe solution, we can `go into Jordan frame' and interpret the scale factor as a scalar field which is coupled to matter. In this case, the redshift of spectral lines of far away sources is no longer interpreted as an effect due to the expansion of the Universe, but due to a growth of coupling constants such that the present transition energies are higher than those in the past which reach us from  far away sources. Hence the Jordan frame physicist does not see an expanding Universe, but growing coupling constants. Nevertheless, the measured redshift of spectral lines is the same for both, the Einstein frame physicist and the Jordan frame physicist.

This difference of the interpretation of the physics at work in the two frames has led to considerable confusion in the literature. In the present paper we want to contribute hopefully not to the confusion but to the elucidation of the issue.
An important point is the fact that one may only consider truly measurable or observable quantities. By laying down a dictionary which translates between background quantities and perturbation variables in both frames, we shall show that the customary density fluctuations or the matter power spectrum are not observables and are actually frame dependent. However, the galaxy number counts which are the true observables, which on small scales, reduce to the density fluctuations and redshift space distortions, are  frame independent.   We shall also show that, a bit like gauge dependence,  frame dependence becomes negligible on small scales.

The rest of this paper is structured as follows. In Section~\ref{s:EJ} we present our dictionary. Most of what we say there can be found in previous literature, see, e.g.~\cite{Catena:2006bd,Chiba:2013mha,Sotiriou:2015lxa,Deruelle:2011, Fujii:2003pa}. For this reason we shall defer most derivations to an Appendix. The original part of this section is our discussion of the scale dependence of the frame effects in different variables. In Section~\ref{s:counts} we apply our findings to the number counts. This section is novel and is the main point of the present work. In Section~\ref{s:con} we conclude.\vspace{0.5 cm} 
\\
\textbf{Conventions and notations:}\\
We work with the $(-,+,+,+)$ convention.\\
The coordinates are ($t,\mathbf{x}$) with $t$ being the conformal time and $\mathbf{x}$ the conformal distance, and $\HH$ the conformal Hubble factor.\\
Quantities in Einstein frame are given without indication, e.g. $\HH$, while quantities in Jordan frame are marked by a tilde, e.g. $\tilde\HH$.\\
The index $0$ means \emph{at the background level}, and is not related to any value today (do this avoids having a too cluttered notation like $\bar{\tilde X}$).\\
The scalar field is denoted as $\phi$ and the Bardeen potentials are $\Psi$ and $\Phi$, while $\varphi$ is the lensing potential.\\
The letter $k$ is used for the kinetic term of the scalar field. In order to avoid confusion, we use $q$ for the momentum.
\newpage

\section{Einstein and Jordan frame - A dictionary}\label{s:EJ}
\subsection{Conformal  relationships}
We want to consider conformally related metrics\footnote{Recall quantities without/with tilde are in Einstein/Jordan frame.},
\be\label{e:ds-rel}
 \mathrm{d}s^2 = F \mathrm{d}\tilde s^2 \,,
\ee
where $F$ is a positive function which may depend on spacetime position.

Length scales in these metrics are related by $\ell = \sqrt{F}\tilde\ell$. This means that a given spacetime interval will measure $\ell$ \emph{units} in Einstein frame, but $\tilde \ell$ \emph{units} in Jordan frame. Using units with $c=\hbar=k_{\rm Boltzmann}=1$, this relation also holds for times while masses and temperatures are related via $m=\tilde m/\sqrt{F}$ and $T=\tilde T/\sqrt{F}$. In these units a mass is given by the inverse of the corresponding Compton wavelength and is also equal to the energy $mc^2$, while a temperature is given by the mean kinetic energy of one degree of freedom at this temperature.

If $F$ is a constant, the two metrics  are related by  a change of units. The important remark then is that measurements are always comparisons, i.e., we only can measure  {\it ratios} between quantities. Usually a measurement is the ratio between the  quantity we are interested in and some reference length scale $\ell_R$ or its inverse.

The interesting case is a dynamical $F$, which may, e.g., depend on a scalar field, $F(\phi)$. In this case, the two conformally related metrics describe the same physics only if we correctly adjust the kinetic term of the scalar field $\phi$ and its coupling to matter. In this case the frame dependence of energies and frequencies, at first sight, seems to imply a frame  dependence of the redshift, the ratio of the frequency a photon emitted at $x_1$ and the one received at $x_2$. The relation between these frequencies in the two frames  is $\nu= \tilde\nu/\sqrt{F}$, so that one might naively infer the redshift
$ (1+z)=\nu(x_1)/\nu(x_2)= \tilde\nu(x_1)\sqrt{F(x_1)}/\left(\tilde\nu(x_2)\sqrt{F(x_2)}\right) = (1+\tilde z)\sqrt{F(x_2)/F(x_2)}$. However, when 'measuring' a frequency at the source we 'compare' it with a standard frequency  or standard length $\ell_R(x_1)$ so that what we truly measure is
\be
 (1+z)=\frac{\nu(x_1)\ell_R(x_1)}{\nu(x_2)\ell_R(x_2)}= \frac{\tilde\nu(x_1)\tilde\ell_R(x_1)}{\tilde\nu(x_2)\tilde\ell_R(x_2)} \,,
 \label{eq:redshift}
 \ee
which is frame {\it independent}. Some more intuitive explanations are given in the appendix. The important point again is that measuring means comparing. Hence the frame dependence of quantities is very subtle and it is sometimes not sufficient to compare dimensionless quantities if they involve measurements at different spacetime points. More details on this point are given in Appendix \ref{app:der}.

We want to describe cosmology with perturbations in both Jordan and Einstein frame. We consider General Relativity with a scalar field and  matter. The actions in the two frames are given by
\be
\SSS= \int \sqrt{-g} \left(\frac{R}{16\pi G} -\frac 12 k(\phi) (\nabla	 \phi)^2 - V+ \LL_{m}(\psi, \phi)\right)\, \mathrm{d}^4x,
\ee
and
\be
\tilde\SSS = \int \sqrt{-\tilde g} \left( \frac{F(\phi) \tilde R}{16\pi G} -\frac 12 \tilde k(\phi) (\tilde\nabla\phi)^2 - \tilde V + \tilde\LL_{m}(\psi, \phi) \right)\, \mathrm{d}^4x.
\ee
Here $\LL_m$ is the matter Lagrangian and $\psi$ stands collectively for all matter fields.
In order to describe the same theory, the variables in Jordan frame (with tilde) and Einstein frame have to be related by
\bea
g_{\mu \nu}  &=&  F \tilde g_{\mu \nu}, \hspace{0.5 cm} \tilde g^{\mu \nu} =  F g^{\mu \nu}, \hspace{0.5 cm} g = F^4 \tilde g,
\label{eq:basic}\\
 V &=& F^{-2}\tilde V, \\
k &=& \frac{3}{16\pi G} \left(\frac{F^\prime}{F}\right)^2 + \frac{\tilde k}{F},    \label{e:non-trivial}
\qquad \mbox{and}\\
 \LL_{m} &=& F^{-2}\tilde\LL_{m}.
\eea
A derivation of the only non-trivial relation, Eq~\eqref{e:non-trivial}, can be found e.g., in~\cite{Wald}.\\  The energy momentum tensor in Jordan frame is given by
\be
\tilde{T}_{\mu \nu} = \frac{2}{\sqrt{\tilde g}}\frac{\partial  (\sqrt{\tilde g}\tilde\LL_{m})}{\partial\tilde g^{\mu \nu}}.
\ee
The quantity in the numerator is frame-independent, while $\sqrt{\tilde g}$ adds a factor of $F^{-2}$ and $\partial \tilde g^{\mu \nu}$ adds a factor of $F$ when converting to Einstein frame so that the whole fraction is multiplied by $F$. The energy-momentum tensors are therefore related as 
\be\label{e:Tmunu}
 \tilde{T}_{\mu \nu} = F T_{\mu \nu}, \hspace{0.5 cm} { \tilde T^\mu}_\nu = F^2 {T^\mu}_\nu, \hspace{0.5 cm} \tilde{T}^{\mu\nu} = F^3 T^{\mu\nu}.
\ee
This is coherent as $\rho \sim T^0_0$ and we then obtain $\tilde\rho = F^2 \rho$, which we expect since $\rho$ has dimension $4$ when counting energy and inverse length dimensions as positive.

\textbf{Important remark:} In what follows, we will define the Jordan frame as the frame where $\ell_R$ is constant and we choose units such that its value is unity. This implies  $\ell_R = \sqrt{F}$ in Einstein frame and $\tilde\ell_R=1$ in Jordan frame. From now on, we will only write $\ell(x)\equiv\sqrt{F}$ to refer to the reference length in  Einstein frame.

\subsection{Background variables}
We now consider a  Friedmann-Lema\^\i tre Universe with metrics
\bea
\mathrm{d}s^2 &=& a^2(-\mathrm{d}t^2+\ga_{ij}\mathrm{d}x^i\mathrm{d}x^j)  \qquad \mbox{and}\\
\mathrm{d}\tilde s^2 &=& \tilde a^2(-\mathrm{d}t^2+\ga_{ij}\mathrm{d}x^i\mathrm{d}x^j)  \qquad \mbox{with } a = \sqrt{F_0} \tilde a 
\eea
in Einstein and Jordan frame respectively. Here $t$ is conformal time and $a$ and $\tilde a$ are the scale factors in Einstein and Jordan frame respectively, while $F_0 \equiv F(\phi_0(t))$, where $\phi_0$ is the background value of the scalar field.  Note that we do not put the subscript $0$ on $V$ and $k$, and the subscript will be also omitted on $\rho$ and $P$ when there is no possible confusion. Moreover, $F_0^\prime \equiv \mathrm{d}F_0/\mathrm{d}\phi_0$ in what follows. Similarly, we define $\ell_0(t)$ as the background value for the reference length. The comoving Hubble parameters are related by
\be
\tilde\HH = \frac{\dot{\tilde a}}{\tilde a}= \frac{1}{{\sqrt{F_0}a}}\frac{\mathrm{d}(\sqrt{F_0}a)}{\mathrm{d}t}=\frac{\dot a}{a}-\frac{\dot{\phi}_0 F_0^\prime}{2 F_0}=\HH -\frac{\dot{\phi}_0 F_0^\prime}{2 F_0} = \HH - \frac{\dot \ell_0}{\ell_0}\,.
\label{eq:hubble_dict}
\ee
The above relation can be understood easily as follows. Let's suppose that the Universe is static in  Jordan frame, hence $\tilde\HH=0$, but that $\dot{\ell}>0$. This means that all lengths appear to be expanding in  Einstein frame, or said differently, that the standard ruler in  Einstein frame is shrinking. In Einstein  frame, the Universe seems to be expanding, hence $\HH >0$.

The matter content in a Friedmann universe is (for symmetry reasons) of the perfect-fluid form, 
	\be
	\left(T{^\mu_\nu}\right) =\mathrm{diag}(-\rho_0, P_0,P_0,P_0),
	\ee
and the same form holds in Jordan frame. Density and pressure  of the two frames are related as 
\be
\rho = F^{-2}\tilde \rho\,, \qquad P = F^{-2}\tilde P \,,
\ee
which is a simple consequence of \eqref{e:Tmunu} (this relationship holds at the background and at the perturbation level).

For  energy momentum `conservation'  we shall assume that the fluid in  Jordan frame does not interact with $\phi$.
This is equivalent to assume that $\tilde\LL_m$ is independent of $\phi$, i.e. $\tilde\LL_m(\psi,\phi)\equiv\tilde\LL_m(\psi)$. In this case Jordan frame fluid satisfies the ordinary  conservation equation,
\be
\nabla^\mu \tilde T\indices{_{\mu \nu}} =0,
\label{eq:consJ}
\ee
while in Einstein frame, the coupling to $\phi$ via the factor $F$ leads to (see e.g.~\cite{Wald})
\be
\nabla^\mu T\indices{_{\mu \nu}} = - \nabla_\nu \phi \frac{F^\prime}{2F} T,
\label{eq:consE}
\ee
where $T$ is the trace of the energy-momentum tensor in Einstein frame. At the background level, as usual, only energy 'conservation' is relevant and  the conservation equations \eqref{eq:consJ} and \eqref{eq:consE} yield
\bea
\dot{\tilde\rho} &=& - \left(\tilde\rho +\tilde P\right)3\tilde \HH, \\
		\dot{\rho} &=& -3 \HH(\rho+P) - \rho \frac{\dot \phi F^\prime}{2F} + P\frac{3\dot \phi F^\prime}{2F}  \,.
\label{eq:rhodotJE}
\eea

The background Einstein equations in  Einstein frame read
\bea
3 \HH^2 &=& 8\pi G_N\left(a^2 \rho + \frac{\dot{\phi_0}^2 k}{2} + a^2 V\right),\\
\HH^2 - 2 \frac{a \ddot{a}}{\dot{a}^2}&=& 8\pi G_N\left(a^2 P + \frac{\dot{\phi_0}^2 k}{2} - a^2 V\right).
\eea
They correspond to the usual Friedmann equations, with the energy density and the pressure of matter and a scalar field, as measured in Einstein frame.

In  Jordan frame, these equations become
\bea
3 \tilde\HH^2 &=& \frac{8\pi G_N}{F_0}\left(\tilde{a}^2 \tilde\rho + \frac{\dot{\phi_0}^2 \tilde k}{2} + \tilde{a}^2 \tilde V\right) - 3 \tilde\HH\frac{\dot \phi_0 F_0^\prime}{F_0},
\label{eq:friedmann_i}\\
\tilde\HH^2 - 2 \frac{\tilde a \ddot{\tilde a}}{\dot{\tilde a}^2}&=& \frac{8\pi G_N}{F_0}\left(\tilde a^2 \tilde P + \frac{\dot{\phi_0}^2 \tilde k}{2} - \tilde a^2 \tilde V\right) + \tilde\HH \frac{\dot \phi_0 F_0^\prime}{F} + \frac{\ddot \phi_0 F_0^\prime}{F_0} + \frac{\dot{\phi_0}^2 F_0^{\prime \prime}}{F_0}.
\label{eq:friedmann_ii}
\eea
We note that Newton's constant appears with a scaling factor $F_0^{-1}$ which is expected, as it has dimension of a length square. The first terms on the right are the density and the pressure as measured in  Jordan frame. Moreover, the additional terms are the signature of the non-minimal coupling between the scalar field and gravity.

It is straightforward to check that these two sets of equations are equivalent if one makes use of the \emph{dictionary}.

\subsection{Perturbations}
We will now consider scalar perturbations at first order. First, note that, if we work in a gauge where the metric is diagonal, this feature will hold in both frames thanks to the conformal relationship \eqref{e:ds-rel}. Hence, we consider only scalar perturbations, since a scalar field only generates scalar perturbations at first order. Therefore vector and tensor perturbations are not affected by a conformal transformation if properly normalized. We use  the Newtonian (or longitudinal) gauge, where the perturbed metrics are given by
	\bea
	\left(g_{\mu \nu}\right) &=& a^2\,  \mathrm{diag}(-1-2\Psi,1-2\Phi,1-2\Phi,1-2\Phi),
	\label{eq:frel_pert}\\
	\left(\tilde g_{\mu \nu}\right) &=& \tilde a^2\,  \mathrm{diag}(-1-2\tilde\Psi,1-2\tilde\Phi,1-2\tilde\Phi,1-2\tilde\Phi),
	\label{eq:frel_pertJ}\,,
	\eea
where $\Phi,~\Psi$ and $\tilde\Phi,~\tilde\Psi$ respectively are the  Bardeen potentials in Einstein and Jordan frames. Second, the scalar field is also perturbed, $\phi(\etac, \mathbf{x}) = \phi_0(\etac) + \delta\phi(\etac, \mathbf{x})$, while the conformal factor is now $F(\etac, \mathbf{x}) = F_0(\etac) + F^\prime \delta \phi(\etac,\mathbf{x})$ (a prime indicates a derivative with respect to the scalar field). We also have for the reference length
\be
\ell(x) = \ell_0(t) + \delta \ell(x) = \sqrt{F_0}\left(1 + \frac{\delta \phi F_0^\prime}{2F_0}\right).
\ee
The \emph{dictionary} for the Bardeen potentials is
	\be
	\Psi = \tilde\Psi + \frac{F_0^\prime \delta \phi}{2F_0} = \tilde\Psi + \frac{\delta \ell}{\ell}, \hspace{0.5 cm}
	\Phi = \tilde\Phi - \frac{F_0^\prime \delta \phi}{2F_0} = \tilde\Phi -  \frac{\delta \ell}{\ell}.
	\label{eq:bardeen_dic}
	\ee
 The denominator containing $\ell$ can be evaluated at the background or at the perturbed level, the difference is of second order. Note that the sum of the Bardeen potentials is frame invariant. This is very satisfactory as they are actually the potential for the Weyl tensor from scalar perturbations which is conformally invariant, see, e.g.~\cite{Durrer:2008aa}. This also means that the lensing potential, which is given in equation (\ref{e:lenspot}), is an integral over the sum of the Bardeen potentials and is frame independent. This confirms the naive expectation that gravitational lensing which describes the deflection of light is conformally invariant.
 \be
 \varphi(\textbf{n},z)=\int_{0}^{r(z)}\mathbf{d}r\frac{r(z)-r}{r(z)r}\left[\Phi(r\textbf{n},t_{now}-r)+\Psi(r\textbf{n},t_{now}-r)\right]
 \label{e:lenspot}
 \ee
 The difference of the Bardeen potentials, however is the anisotropic stress tensor which is not frame invariant.
 
 Note that even though the Bardeen potentials are gauge invariant, i.e. invariant under linearized coordinate transformation, they are not frame invariant and therefore not directly observable.
 
Let us now turn to the energy-momentum tensors. For simplicity we consider only perfect fluid matter, without anisotropic stress, such that the perturbed tensors are of the form
	\bea
	T\indices{^\mu_\nu} &=&(\rho_0 + \delta\rho)u^\mu u_\nu + (P_0 + \delta P)\left(u^\mu u_\nu + {\de^\mu}_\nu\right) \,,\\
	\tilde	T\indices{^\mu_\nu} &=&(\tilde\rho_0 + \delta\tilde\rho)\tilde u^\mu \tilde u_\nu + (\tilde P_0 + \delta\tilde P)\left(\tilde u^\mu\tilde u_\nu + {\de^\mu}_\nu\right) \,.
	\eea
The \emph{dictionary} relating the perturbations in both frames is given by
	\begin{align}
	\delta(x) &= \frac{\rho(x) - \rho_0(t)}{\rho_0(t)}  
	=  \frac{F^2 \tilde \rho(x) - F_0^{-2} \tilde\rho_0(t)}{F_0^{-2} \tilde\rho_0(t)}\\
	&= \frac{F_0^{-2} \tilde \rho(x) \left( 1 - 2 \frac{F_0^\prime \delta \phi}{F_0} \right) - F_0^{-2} \tilde\rho(t) }{F_0^{-2} \tilde\rho(t)} \\
	&=\tilde\delta(x)- 2 \delta\phi \frac{F_0^\prime}{F_0} 
	= \tilde\delta(x) - 4 \frac{\delta \ell}{\ell}\, .
	\label{eq:delta_rho}
	\end{align}
The same holds for the pressure perturbation, namely
\begin{align}
	\frac{\delta P(x)}{P_0}  &= \frac{\delta\tilde P(x)}{\tilde P_0} - 4 \frac{\delta \ell}{\ell}\,.
	\label{eq:delta_P}
	\end{align}
Eq.~\eqref{eq:delta_rho} can be understood as follows: if $\tilde\delta = 0$ (Jordan frame), then matter is distributed  uniformly on a given time slice. However, at a given spatial position, $\ell$ may be slightly larger than on the rest of the time slice (then the standard ruler is smaller). Lengths appear larger, and hence energies smaller. The minus sign is then in agreement with the fact that, at this particular position, the energy density will appear smaller in Einstein frame.

Let us finally express the velocity of the fluid. This needs some care. We use that the velocity is normalized in either frame, $\tilde u^2 =  \tilde u^\mu \tilde u^\nu \tilde g_{\mu \nu} = u^2= u^\mu  u^\nu g_{\mu \nu}=-1$, so that one has
\be
\tilde u^\mu = \sqrt{F}u^{\mu} = \ell u^\mu = \frac{1}{\tilde a}\left(1-\tilde\Psi,\tilde\bv\right),
\ee
Here the $\tilde u^0$ term is fixed by the normalization and we have defined the velocity perturbation in Jordan frame, $\tilde\bv$.
 In Einstein frame we obtain
\begin{align}
u^\mu &= \ell^{-1} \tilde u^\mu  \\
&= \ell_0^{-1}\left(1 - \frac{\delta \ell}{\ell}\right) \tilde a^{-1} (1-\tilde\Psi , \tilde{\mathbf{v}}) \\
&= a^{-1} \left(1 - \left(\tilde\Psi + \frac{\delta \ell}{\ell} \right)  , \tilde{\mathbf{v}}\right) ~=~  a^{-1} (1-\Psi , {\mathbf{v}}).
\label{eq:vit_pert}
\end{align}
where we have used equation \eqref{eq:bardeen_dic} relating $\Psi$ and $\tilde\Psi$ and the fact that $\mathbf v$ is already first order. Hence the peculiar velocity is not modified (at first order), which is quite intuitive for a dimensionless local quantity.

We also want to compute the redshift perturbation in Einstein frame. This is possible by taking the usual formula, see e.g.~\cite{Bonvin:2011bg}, and taking into account that the {\it measured} frequency is $(k\cdot u)\ell$ we obtain in longitudinal gauge
\be
\delta z(\mathbf{n}, z_0) = -(1+z_0)\left( \Psi+ \mathbf{n}\cdot \mathbf{v} + \int^{r(z_0)}_0 \mathrm{d}r(\dot{\Psi} +\dot\Phi)\;  - \frac{\delta \ell}{\ell}\right)\,,
	\label{eq:deltazEfor}
\ee
where $r(z)$ denotes the conformal distance at the the background redshift $z_0$. 
Here all the terms are evaluated at emission, $(t(z_0), r(z_0)\bn) $ (we assume the observer to be situated at $\mathbf{x}=0$).
  As usual,  we omit terms at the observer which add only a monopole or a dipole contribution to the final results. Note that the terms with $\Psi$ and $\delta \phi$ add up, and accordingly to the relation between the Bardeen potentials in both frames \eqref{eq:bardeen_dic}, they simply give $\tilde\Psi$ which makes this relation frame invariant. The sum in the integral is, as mentioned, frame invariant, hence the redshift perturbation is frame independent. Interestingly it is not gauge invariant, see~\cite{Bonvin:2011bg}. This comes from the fact that the split into $z = z_0+\de z$ depends on the time slice i.e. on the chosen gauge.

The density perturbation and therefore also the matter power spectrum, $P(q)=|\de(q)|^2$ is frame dependent. Note also that in cosmology $q$ is the comoving wave number which is frame independent, hence Fourier transforms do not affect the frame dependence of first order perturbation variables.

\subsection{Perturbed conservation equations}
Let us determine the perturbed `conservation' equations. We have seen that already at the background level, matter in Einstein frame is interacting with the scalar field and  $\nabla^\mu{T_\mu}^\nu \neq 0$. We now  perturb these equations at first order (see appendix for more details). Here, we follow the idea presented in~\cite{Sotiriou:2015lxa}.

In Einstein frame, the conservation equations read\footnote{We use the usual notation $P_0=\omega \rho_0$ and $\delta P = c_s^2 \delta \rho$, with $\omega$ and $c_s^2$ frame invariant.}
\bea
\hspace{-0.21cm}    \dot\delta + 3\HH\left(c_s^2-\omega\right) \delta +(1+\omega) \mathbf\nabla \cdot \mathbf v - 3  (1+\omega) \dot \Phi &=&  (3\omega-1) \left(\frac{F_0^\prime}{2F_0}  \delta \phi \right)^\bullet + 
    \frac{3 F_0^\prime \dot \phi}{2F_0}\left(c_s^2 - \omega\right)\delta \nonumber \\
    &=&  (3\omega-1) \left(\frac{\delta \ell}{\ell}\right)^\bullet + 
   3 \frac{\dot{\ell}}{\ell}\left(c_s^2 - \omega\right)\delta, \\
 \hspace{-2.5cm}     \dot{\mathbf v} + \HH  (1-3 \omega)\mathbf v + \mathbf \bnabla \Psi + \frac{c_s^2}{(1+\omega)}\nabla \delta&=& \frac{1-3\omega}{1+\omega}\frac{\bnabla \delta\ell}{\ell_0} 
   +(1-3\omega) \frac{\dot \ell_0}{\ell_0} \mathbf v.
\eea
Considering the same equations in Jordan frame, on can use the various equations in the \emph{dictionary} to obtain the usual relations
\bea
    \dot{\tilde\delta} + 3\tilde \HH \tilde \delta \left(c_s^2-\omega\right) + \mathbf\nabla \cdot \mathbf {v} - 3 \dot {\tilde\Phi} (1+\omega)&=& 0, \\
    \dot{\mathbf v} + \tilde\HH \mathbf v (1-3\omega) + \mathbf \nabla \tilde\Psi + c_s^2 \frac{\mathbf{\nabla} \tilde{\delta}}{(1+\omega)} &=& 0.
\eea

Note that, as expected, these equations are the usual ones without coupling to the scalar field $\phi$, see e.g.~\cite{Durrer:2008aa}. In Einstein frame, the right hand side is the effect of the coupling between matter and the scalar field. In the second equation, they can be interpreted as the \emph{fifth force}, which refers to any force modifying the usual geodesic equation. In the case of a Universe filled with radiation only ($\omega = c_s^2 = 1/3$), the equations are not modified. This is simply a consequence of the fact that massless particles are invariant under conformal transformations.

\subsection{Perturbed Einstein equations}
Regarding the perturbations, in Einstein frame, one has $\Phi = \Psi$ if the anisotropic stress can be neglected. From now on, we will also assume vanishing pressure, since we are mainly interested in a $\La$CDM Universe with a scalar field at redshifts $0<z<6$ where the fluid matter is pressureless \footnote{Adding it does not lead to any principal difficulties, but the equations become just more cumbersome.}. The perturbed Poisson equation ($00$ constraint) is\footnote{As usual for scalar perturbations, we introduce the velocity potential $v$ with $\mathbf v = \mathbf \nabla v$.}
\bea
\Delta \Phi 
- 4 \pi G a^2 \rho \left(\delta -3 \HH v\right)
=
4\pi G 
\left(
    \left(
        3\HH k \dot \phi 
        + \frac{k^\prime}{2} \dot{\phi}^2
        + \frac{a^2 V^\prime}{2}
    \right) \delta \phi
    +  k \dot\phi \dot{\delta \phi}
    - k \dot{\phi}^2 \Phi
\right)
\label{eq:poissonE}
\eea
The perturbed scalar $(0i)$ constraint is
\bea
\dot \Phi + \HH \Phi + 4 \pi G a^2 \rho v=4\pi G k \dot \phi  \delta\phi.
\eea
The scalar dynamical equation is
\bea
\ddot\Phi
+ 3 \HH \dot\Phi
- \HH^2 \Phi 
+ 2 \frac{\ddot a}{a}\Phi 
=
4\pi G 
\left( 
    \frac 12 k^\prime \dot{\phi}^2 \delta \phi 
    - a^2 V^\prime \delta \phi 
    +k \dot \phi \dot{\delta \phi} 
    - k \dot{\phi}^2 \Phi 
\right).
\label{eq:efe_pert_ef_ii}
\eea
In Jordan frame, the equations are quite long. We give the full expressions in Appendix~\ref{a:eeJordan}. Here we just state that the off diagonal part of the dynamical equations in Jordan frame yield
\be
\tilde\Phi-\tilde\Psi =2\frac{\de\ell}{\ell}\,,
\ee
which is in agreement with Eqs.~\eqref{eq:bardeen_dic}.

\subsection{Scaling of the frame dependence}
In this section, we investigate the frame dependence of different quantities depends on the considered wavelength. As shown by the dictionary for the density perturbation \eqref{eq:delta_rho}, the difference is given by the quantity $\delta \ell / \ell$. We estimate its order of magnitude. We will assume that we work on  subhorizon scales such that $q \gg \HH$, and for example $(q^2 + \HH^2)\Phi \sim q^2 \Phi$. The power spectrum of the matter perturbation is $\mathcal{P}(q) \sim \delta^2$. Hence, the power spectra $\mathcal{P}, \tilde{\mathcal{P}}$ in both frames are related as
\be
\tilde{\mathcal{P}} \sim \mathcal{P} + 8 \delta\cdot  \frac{\delta \ell}{\ell} + 16 \left( \frac{\delta \ell}{\ell} \right)^2.
\label{eq:powspec}
\ee
In order to evaluate the last terms, we will use the Einstein equations and make several assumptions. We work now in Einstein frame. We have at our disposal two scales: a time/length scale given by $\HH^{-1}$, and a momentum given by $q$. Assuming there is no other scale governing the evolution of the scalar field, we can estimate that
\be
\dot \phi \sim \HH \phi, \hspace{0.5 cm} \dot{\delta\phi} \sim \HH \delta\phi.
\label{eq:pow_spec}
\ee 
Moreover, we will suppose that $F$ is a polynomial function in $\phi$ such that $F^\prime \sim F/\phi$, and the same for $k$ (the coefficient of the kinetic term).\\
Taking the dynamical perturbed Einstein's equation \eqref{eq:efe_pert_ef_ii}, we can estimate
\be
\HH^2 \Phi \sim \HH^2 \frac{\delta \phi}{\phi}, 
\label{eq:scale}
\ee
where we used the first Friedmann equation \eqref{eq:friedmann_i} to evaluate $4\pi Gk \dot{\phi}^2 \sim 4\pi Ga^2 V \sim \HH^2$.\\
The conclusion is that the relative perturbation of the scalar field scales roughly as the Bardeen potentials
\be
\Phi \sim \frac{\delta \phi}{\phi}.
\label{eq:scale_ii}
\ee
This implies that the difference between the two density perturbations, $\delta$ and $\tilde \delta$ scales in the same way:
\be
\frac{\delta \ell}{\ell} \sim \frac{\delta \phi F^\prime}{F} \sim \frac{\delta \phi}{F} \frac{F}{\phi} \sim \Phi.
\ee
Performing the same order of magnitude evaluation on the perturbed Poisson equation \eqref{eq:poissonE}, we can estimate the relation between the Bardeen potential and the density perturbations on subhorizon scales, 
\be 
q^2 \Phi \sim \HH^2 \delta.
\ee 
By combining the three previous relationships, we find
\be
\frac{\delta \ell}{\ell} \sim \left(\frac{\HH}{q}\right)^2 \delta.
\ee
If we now turn our attention to the relation between the two power spectra in equation \eqref{eq:pow_spec}, we obtain the order of magnitude relation
\be
\tilde{\mathcal{P}} \sim \mathcal{P} 
\left(
    1 + \mathcal{O}\left(\left[\frac{\HH}{q}\right]^2\right)
\right).
\label{eq:pow_spec_diff}
\ee
This shows that, even if the power spectrum is frame dependent, hence not an observable, the difference at small scales is negligible. In galaxy surveys, on small, widely subhorizon scales frame effects can therefore be neglected. It is also on these scales that thew number counts can be expressed in terms of the power spectrum. On large scales, $q\sim \HH$, however frame effects are as important as other relativistic effects. We shall now show that, as it must be for a true observable, in the galaxy number counts all frame effects cancel.

\section{Cosmological number counts}\label{s:counts}
\subsection{Number count in Jordan frame}
In this section, we will follow the notation of \cite{Bonvin:2011bg}. See Appendix for more details. In  Jordan frame, as the value of $\ell$ is constant, the same formula holds, namely
\be
\tilde \Delta(\mathbf{n},z)  = \tilde \delta_z(\mathbf{n},z) +  \frac{\delta \tilde V(\mathbf{n},z)}{\tilde V(z)}
\label{eq:countj}
\ee
Here $\tilde \Delta$ is the number count, which is actually observed in galaxy survey, $\delta_z$ is the density perturbation \textbf{in redshift space}, and is \textbf{not} the usual density perturbation of the energy-momentum tensor and $\delta V$ is the volume perturbation. We will show that both these terms are not frame invariant, but all the frame dependent terms  cancel, and we shall find that  the number counts are frame invariant.

\subsection{Number counts in Einstein frame}
The main issue in Einstein frame is the variation of the ruler. Recall that, in this frame, we have a function $\ell(x)$ defining the size of our ruler at a given spacetime point. The computations need to be modified to take this into account. Regarding  density perturbations in redshift space, we need to take into account that density is related to $N$ and $V$ by $\rho=mN/V$. Here, for simplicity we assume a Universe made out of particles (galaxies) of fixed mass $m$ of which we find $N$ in the volume $V$.   In Jordan frame, or in the usual formulation of General Relativity, this mass is constant and it cancels in all expressions. However, this is not true in our Einstein frame where the ruler is not constant\footnote{Note that for dimensional reasons, one should actually take the ratio with the reference length in Jordan frame $l_J$. As this value is constant, we can simply omit it, because it  cancels in all the ratios.}. We can then compute the redshift density perturbations taking into account that the mass scales like $m\propto \ell^{-1}$:
\begin{align}
\delta_z(\mathbf{n},z) 
&= 
\frac{\rho(\mathbf{n},z)- \rho_0(z)}{\rho_0(z)} 
=
\frac{
	\frac{N(\mathbf{n},z) \ell(\mathbf{n},z)^{-1}}{V(\mathbf{n},z)} -
	\frac{N_0(z) \ell_{0}(z)^{-1}}{V_0(z)} 
}
{	\frac{N_0(z) \ell_{0}(z)^{-1}}{V_0(z)}  } \\
&=
\Delta(\mathbf{n},z) +  \frac{\delta z}{\ell_{0}} \frac{\mathrm{d}\ell_{0}}{\mathrm{d}z_0} - \frac{\delta \ell}{\ell_{0}} - \frac{\delta V}{V_0}.
\label{eq:deltazE}
\end{align}
More details are given in Appendix~\ref{app:der}. The important point is that, in the second equality, we use that $\rho = m N/ V$, and the mass in Einstein frame scales as $\ell^{-1}$, hence the appearance of this factor. Moreover, note that here we want to isolate the number count, hence we only need to convert the masses $m$ to the number of galaxies $N$, which brings exactly one factor of $\ell^{-1}$. The volume $V$ is still measured in  Einstein frame, hence we do not have factors of $\ell^3$ appearing which would convert it to Jordan frame.

The formula for the number counts in Einstein frame is then given by 
	\be
		\Delta(\mathbf{n},z)  = \delta_z(\mathbf{n},z)
		+  \frac{\delta V(\mathbf{n},z)}{V(z)} 
		- \frac{\delta z}{\ell_{0}} \frac{\mathrm{d}\ell_{0}}{\mathrm{d}z_0}
		 + \frac{\delta \ell}{\ell_{0}}.
		\label{eq:counte}
	\ee
To relate it to the number count in  Jordan frame, we need to relate the corresponding density perturbation in the redshift space, $\de_z$ and the volume fluctuation $\de V$. We can perform this computation in two ways.

We first present a naive and quick version, before giving the detailed derivation in the next two sections. Consider a physical quantity $f$ in Einstein frame whose energy dimension is $n$. We can relate the perturbation in Einstein and in Jordan frame as (see derivation in Appendix \ref{app:der})
\begin{align}
    \frac{\delta f}{f} 
    \equiv  
    \frac{\delta f(\mathbf{n},z)}{ f_0(z)} 
    = 
    \frac{\delta \tilde f}{\tilde f}  
    + n \frac{\delta z}{\ell_{0}}\frac{\mathrm{d} \ell_{0}}{\mathrm{d} z_0} 
    - n \frac{\delta \ell}{\ell}.
\label{eq:generaletoj}
\end{align}

We know that $\delta \rho$ and $V$ have energy dimensions $4$ and $-3$ respectively. Hence, their sum for the number count in equation \eqref{eq:counte}  gives 
\be
    \delta_z(\mathbf{n},z)
    +  \frac{\delta V(\mathbf{n},z)}{V(z)} 
    =
    \tilde \Delta(\mathbf{n},z) 
    + \frac{\delta z}{\ell_{0}}\frac{\mathrm{d} \ell_{0}}{\mathrm{d} z_0} 
    - \frac{\delta \ell}{\ell_{0}},
\label{eq:corr}
\ee
where we have used the definition of $\tilde \Delta$ \eqref{eq:countj}. This  precisely cancels the remaining last term of equation \eqref{eq:counte} .

This derivation is somewhat dangerous as e.g. conformal time $t$ and distance $r$ even though they usually do have dimensions do not transform in this way while the dimensionless Bardeen potentials do transform.
However, considering the the metrics \eqref{eq:frel_pert} and \eqref{eq:frel_pertJ} we realize that $\Psi$ and $\tilde\Psi$ correspond to changes in physical time intervals while $\Phi$ and $\tilde\Phi$ correspond
to changes in the inverse physical distance (at first order) this explains their behavior under conformal transformations given in \eqref{eq:bardeen_dic}. Because of this subtlety we now present a more formal derivation of the same result.

\subsubsection{Density perturbation}
We want to show how we can obtain the relation \eqref{eq:generaletoj}, but this time by following the approach of \cite{Bonvin:2011bg}. As shown, the density in redshift space is given by
\be
\delta_z(\mathbf{n},z) = \frac{\delta \rho (\mathbf{n},z)}{\bar \rho(z_0)} - \frac{\mathrm{d}\rho_0}{\mathrm{d}z_0} \frac{\delta z (\mathbf{n},z)}{\rho_0(z_0)}.
\label{eq:deltazEdef}
\ee
The first term is simply the relative energy perturbation given by the dictionary equation for $\delta$ \eqref{eq:delta_rho}.

The second term of \eqref{eq:deltazEdef} is more tricky. As both $z$ and $z_0$ are frame independent, $\delta z$ must to be so as well. Considering the definition in \eqref{eq:redshift} including the ruler-dependence, the redshift in  Einstein term picks up a factor $\ell$. The second term can be evaluated  using the conservation equation \eqref{eq:rhodotJE} (with $P=0$) and the time derivative of $z_0$. This last quantity is obtained directly from the definition of $z$, \eqref{eq:redshift} and we find\footnote{Recall that both $a$ \textbf{and} $\ell$ are time dependent.}
	\be
	\frac{\mathrm{d}z_0}{\mathrm{d}t} = -(1+z_0) \left(\HH - \frac{F_0^\prime \dot{\phi}_0}{2F_0} \right) = -(1+z_0) \tilde \HH.
	\label{eq:dzdtE}
	\ee 
The last equality uses the dictionary between $\HH$ and $\tilde\HH$. As $z_0$ and $t$ are frame-invariant, so is this derivative. 
Combining the previous results yields 
	\be
	\delta_z = \tilde \delta_z - 4\frac{\delta \ell}{\ell_{0}} + 4 \frac{\mathrm{d}\ell_{0}}{\mathrm{d}z_0} \frac{\delta z}{\ell_{0}},
	\label{eq:dzEfinal}
	\ee
where we use the chain rule to go from one derivative to another (recall $\ell_0 = \sqrt{F_0}$), and
\be
\tilde \delta_z = \tilde \delta - \frac{3 \delta z}{1 + z_0}.
\ee
Note that equation \eqref{eq:dzEfinal} is in agreement with \eqref{eq:generaletoj}, with $n=4$.

\subsubsection{Volume perturbation}
We will briefly present the relation of the volume perturbations and mention which corrections need to be taken into account in Einstein frame. The most important point is that, in this frame, we do \emph{not} have that $a^{-1} = 1+z_0$ (where $a$ is the scale factor at the emission). Recall the time derivative of $z_0$ given by equation \eqref{eq:dzdtE}. Then, equation (14) of \cite{Bonvin:2011bg} contains a prefactor of the form
\be
\frac{a^3}{1+z} \frac{1}{\HH - \frac{\dot{\phi}_0 F_0^{\prime}}{2F_0}} = \frac{\ell^3}{(1+z)^4} \frac{1}{\HH - \frac{\dot{\phi}_0 F_0^\prime}{2F_0}}.
\ee
The remaining steps are as in Ref.~\cite{Bonvin:2011bg}, provided we include this correction. Hence we find
\be
\frac{\de\tilde {V}}{\tilde V_0} = \frac{\de {V}}{ V_0}+  3\frac{\delta \ell}{\ell_{0}} -3 \frac{\mathrm{d}\ell_{0}}{\mathrm{d}z_0} \frac{\delta z}{\ell_{0}}\,.
\ee
Note that the factor $\ell^3$  exactly brings the factors predicted by equation \eqref{eq:generaletoj} with $n=3$.

\section{Conclusions}\label{s:con}
In this paper we have studied the frame dependence of cosmological perturbation variables. We have shown that gauge invariance does not guarantee frame invariance (consider, e.g., the Bardeen potentials) and that, on the other hand, there are gauge dependent quantities (e.g. redshift, velocities) which are frame invariant. We have finally shown that the physical, observable number counts are frame invariant, while the density power spectrum is not. This remains true when adding redshift space distortions since velocities are frame independent. However, the frame dependence is relevant only on large scales, comparable to the Hubble scale,  where density fluctuations are not directly observable but acquire relativistic corrections.
We have also shown that the lensing potential, which is a weighted integral of the sum of the Bardeen potentials is frame independent.

We summarize the gauge and frame dependence of the interesting variables in the final table below. While we do not spell out the gauge dependence which can be found e.g. in~\cite{Durrer:2008aa}, we explicitly give the relation between the corresponding quantities in Einstein and Jordan frame.

\begin{table*}[ht]
\caption{Gauge and frame dependence of various quantities.  ~}      \label{TableClass}     
\begin{center}                         
    \begin{tabular}{|c|c|c|c|}
    \hline\hline  & & \multicolumn{2}{l|}{} \\
    Quantity            & Gauge dependent & \multicolumn{2}{c|}{Frame dependent~~~~~~} \\  \hline
    {\bf Background} &&& \\
    Density $\rho_0$ & Yes  &Yes & $\rho_0 = F_0^{-2}\tilde \rho_0$\\
    Pressure $P_0$ &Yes & Yes & $P_0 = F_0^{-2}\tilde P_0$\\
    Redshift $z_0$  & Yes  & No & \\
    Observed redshift $z$   & No & No & \\ \hline
    {\bf Perturbations} &&& \\
    Density $\delta$ & Yes & Yes &  $\delta=\tilde\de -4\de\ell/\ell$ \\
    Velocity $\bv$ & Yes & No & \\
    Bardeen pot. $\Psi$ & No & Yes & $\Psi=\tilde\Psi +\de\ell/\ell$\\
    Bardeen pot. $\Phi$ & No & Yes & $\Phi=\tilde\Phi -\de\ell/\ell$\\
    Lensing potential $\varphi$ & No & No & \\
    Redshift density $\delta_z$ & No & Yes & $\delta_z=\tilde\de_z - 4\frac{\delta \ell}{\ell} + 4 \frac{\mathrm{d}\ell}{\mathrm{d}z} \frac{\delta z}{\ell}$ \\
    Volume perturbation $\delta V/V$ & No & Yes & $\de V/V=\de\tilde V/\tilde V +3\frac{\delta \ell}{\ell} -3\frac{\mathrm{d}\ell}{\mathrm{d}z} \frac{\delta z}{\ell}$\\
    Number counts $\De(\bn,z)$ & No & No &\\  &&& \\
    \hline
    \end{tabular}
    \end{center}
\end{table*}

\section*{Acknowledgement}
We thank Camille Bonvin for helpful discussions.
This work is supported by the Swiss National Science Foundation.

\appendix
\section{Some explanations}
We give here further intuitive explanations for some of the equations of the paper.

Let us recall the new definition of the redshift given by \eqref{eq:redshift}. We consider a photon emitted at position $e$ and received at position $r$:
\be
    (1+z)=\frac{\nu_e\ell_e}{\nu_r\ell_r}=\frac{\lambda_r\ell_e}{\lambda_e\ell_r} \,.
    \label{eq:redshift_ii}
\ee
Here $\ell$ represents the measured length of a given, fixed physical process, i.e. our ruler.
Let's consider two examples.
\begin{itemize}
    \item First, suppose that $\lambda_r = \lambda_e$ but $\ell_e > \ell_r$. With the usual definition of $z$, this would mean there is no redshift, because both wavelengths are measured equal. However, in the framework of conformal frames, we have
    \be 1+z = \frac{\ell_e}{\ell_r}\ee
    which implies that $z>0$. This is due to the fact that the ruler which is used is expanding with the considered wave, hence its wavelentgh is seen to be constant. Because the ruler is expanding, the reference length, which is itself not being stretched, appears smaller and smaller hence $\ell_e > \ell_r$.
    \item On the other hand, one can consider a process where $\lambda_e < \lambda_r$ and $\ell_e < \ell_r$ such that both ratios cancel and give $z=0$. This means that the ruler is expanding, hence the measured reference length appears smaller, but the wavelength of the travelling photon also appears to be decreasing. In fact, nothing is being redshifted, the only modification is that the ruler is being stretched.
\end{itemize}
The second point illustrates the following interesting fact: If the physical process which is used to define $\ell$ is also undergoing the same `redshift', we would naively think that $z=0$. This is why it is important that the comparisons are made \emph{locally}: at each spacetime point, one has to reproduce the reference physical process in their laboratory, and calibrate in this way the values for $\ell(x)$.

Let us now turn our attention to the conservation equations in  Einstein frame \eqref{eq:rhodotJE}. Consider a box filled with matter only. We assume for simplicity that $\tilde \HH=0$ in Jordan frame, and suppose that $\dot{\ell}>0$, which means that the standard ruler in  Einstein frame is shrinking. We can have in mind a fixed box of a given size, and inside a ruler that is getting smaller over time.

By using the \emph{dictionary} for the Hubble factors \eqref{eq:hubble_dict}, we find $\HH = (\dot \phi F^\prime)/(2F) =\dot{\ell}/\ell$ and equation \eqref{eq:rhodotJE} reads
\be
\dot \rho = -\rho\left(\frac{3 \dot{\ell}}{\ell}  + \frac{\dot \ell}{\ell}\right) - P \left( \frac{3 \dot{\ell}}{\ell} - \frac{3 \dot{\ell}}{\ell} \right) =  -\rho\left(\frac{3 \dot{\ell}}{\ell}  + \frac{\dot \ell}{\ell}\right) < 0
\label{eq:rhodot_ex}
\ee
First, we note that we have two contributions: the first one is the usual term corresponding to the (apparent) volume expansion of the box, while the second term corresponds to the (apparent) decreasing masses: $\dot \ell >0$ and $m \propto \ell^{-1}$, the masses appear to get smaller in Einstein frame. Regarding the pressure, it is interesting to note that both terms cancel. This should not be a surprise, since the contribution of the pressure to the variation of the energy is given by $W = - P \delta V$. But in this case, the true volume (the volume of the box, measured in Jordan frame) is not varying.

\section{Perturbed Einstein equations in Jordan frame}\label{a:eeJordan}
In  Jordan frame, the perturbed Poisson equation ($00$ eqn.) reads
\bea
\Delta\tilde\Phi 
- \frac{4 \pi G \tilde{a}^2 \rho}{F} 
\left(\tilde \delta -3 \tilde \HH v\right)
&=&
\frac{4\pi G}{F_0} 
\left(
    \left(
        3\tilde \HH \tilde k \dot \phi 
        + \frac{\tilde k^\prime}{2} \dot{\phi}^2
        + \frac{a^2 \tilde V^\prime}{2}
    \right) \delta \phi
    + \frac{3}{2} \tilde k \dot \phi \dot{\delta \phi}
    - \tilde k \dot{\phi}^2 \tilde \Phi
\right) \nonumber \\
&-&
\frac{4\pi G}{F_0}
\left(
    \tilde{a}^2 \tilde \rho_0 \frac{F_0^\prime}{F_0}\delta\phi
    - \frac{\tilde k}{2}\dot{\phi}^2 F_0^\prime \delta\phi
    + \tilde{a}^2 \tilde V \frac{F_0^\prime}{F_0} \delta \phi
\right) \nonumber \\
&+& 
\frac{F_0^\prime}{2F_0}
\left(
    \Delta \delta \phi
    - 3 \tilde \HH^2 \delta \phi 
    + 3 \tilde \HH \dot{\phi} \tilde \Phi 
    + 3 \phi\dot{\tilde{\Phi}}
\right)+
\frac{\dot{\phi}F^{\prime \prime}_0}{2F_0} \dot{\delta \phi}.
\label{eq:poissonJ}
\eea
The perturbed scalar $(0i)$ constraint is
\bea
\dot{\tilde \Phi} 
+ \tilde \HH \tilde \Phi 
+ \frac{4 \pi G \tilde{a}^2 \tilde \rho}{F_0} v
&=&
\frac{4\pi G k \dot \phi}{F_0}\delta\phi +
\frac{F^\prime_0}{2F_0}
\left(
    \tilde \HH \delta \phi 
    + \frac{F_0^\prime}{F_0}\dot{\phi} \delta \phi 
    -\dot\phi \tilde \Phi  
    + \dot{\delta \phi} 
\right)
+\frac{F^{\prime\prime}_0}{2F_0}\dot{\phi} \delta \phi.~~~ 
\eea
The trace of the dynamical equation is 
\bea
\ddot \Phi + 3 \tilde\HH \dot\Phi  -\tilde\HH^2 \Phi + 2 \frac{\ddot a}{a}\Phi 
&=&
\frac{4\pi G}{F_0} 
\left(
    \frac{k}{2} \dot \phi \dot{\delta \phi} - k \dot{\phi}^2 \Phi - a^2 V^\prime \delta \phi + \frac 12 k^\prime \dot{\phi}^2 \delta \phi
\right) \nonumber \\
&+&
\frac{4\pi G}{F_0} 
\left(
    \tilde a^2 \tilde V \left(\frac{F_0^\prime}{F_0}\right)^2 \delta \phi
    +\frac{\tilde k}{2}\dot{\phi}^2 \frac{F_0^\prime}{F_0^2}
\right) \nonumber \\
&-&
\frac{F_0^\prime}{2F_0}
\left(
    2 \tilde\HH^2 
    - 4\frac{\ddot a}{a}
    + \tilde \HH \frac{F_0^\prime}{F_0}\dot \phi 
    + \left(\frac{F_0^\prime}{F_0}\right)^2\dot{\phi}^2
\right) \delta \phi \nonumber \\
&+&
\frac{F_0^\prime}{2F_0}
\left(
    3\tilde \HH + \dot \phi
\right) \dot{\delta \phi} + \frac{F_0^\prime}{2F_0} \ddot{\delta \phi} \nonumber \\
&-&
\frac{F_0^\prime}{2F_0}
\left(
    2 \tilde \HH \dot \phi \Phi 
    + 2 \ddot{\phi} \Phi
    + 3 \dot \phi \dot \Phi
\right) \nonumber \\
&+&
\frac{F_0^{\prime\prime}}{2F_0}
\left(
    3 \tilde \HH \dot \phi \delta \phi
    + \ddot \phi \delta \phi
    + \frac{F_0^\prime}{F_0} \dot{\phi}^2 \delta \phi
    + \ddot \phi \dot \delta \phi
    - \dot{\phi}^2 \Phi
\right) \nonumber \\
&+& \frac{F_0^{(3)}}{2F_0} \dot{\phi}^2 \delta \phi\,,
\label{eq:efe_pert_jf_ii}
\eea
while its traceless part simply yields
\be
\tilde\Phi-\tilde\Psi =2\frac{\de\ell}{\ell}\,.
\ee

\section{Some derivations \label{app:der}}
We present the derivation of equation \eqref{eq:generaletoj}. Consider a physical quantity $f$ with energy dimension $n$.

\begin{align}
\frac{\delta f}{f} &\equiv  \frac{\delta f(\mathbf{n},z)}{ f_0(z)} = \frac{f(\mathbf{n},z) - f_0(z)}{f_0(z)} \\
			&= 
\frac{\tilde f(\mathbf{n},z) \ell(\mathbf{n},z)^{-n}- \tilde f_0(z) \ell_{0}(z)^{-n}}{ f_0(z)\ell_{0}(z)^{-n}}  \\
			&=
\frac{ \tilde f(\mathbf{n},z) \ell_{0}(z)^{-n}\left( 1 + n \frac{\delta z}{\ell_0}\frac{\mathrm{d} \ell_{0}}{\mathrm{d} z_0} - n \frac{\delta \ell}{\ell} \right)- \tilde f_0(z) \ell_{0}(z)^{-n}}{ f_0(z)  \ell_{0}(z)^{-n} }  \\
			&= 
\frac{\delta \tilde f}{\tilde f}  + n \frac{\delta z}{\ell_0}\frac{\mathrm{d} \ell_{0}}{\mathrm{d} z_0} - n \frac{\delta \ell}{\ell}.
			\label{eq:generaletoj_proof}
\end{align}

\begin{figure}[h!]
\centering
	\includegraphics[width=9cm]{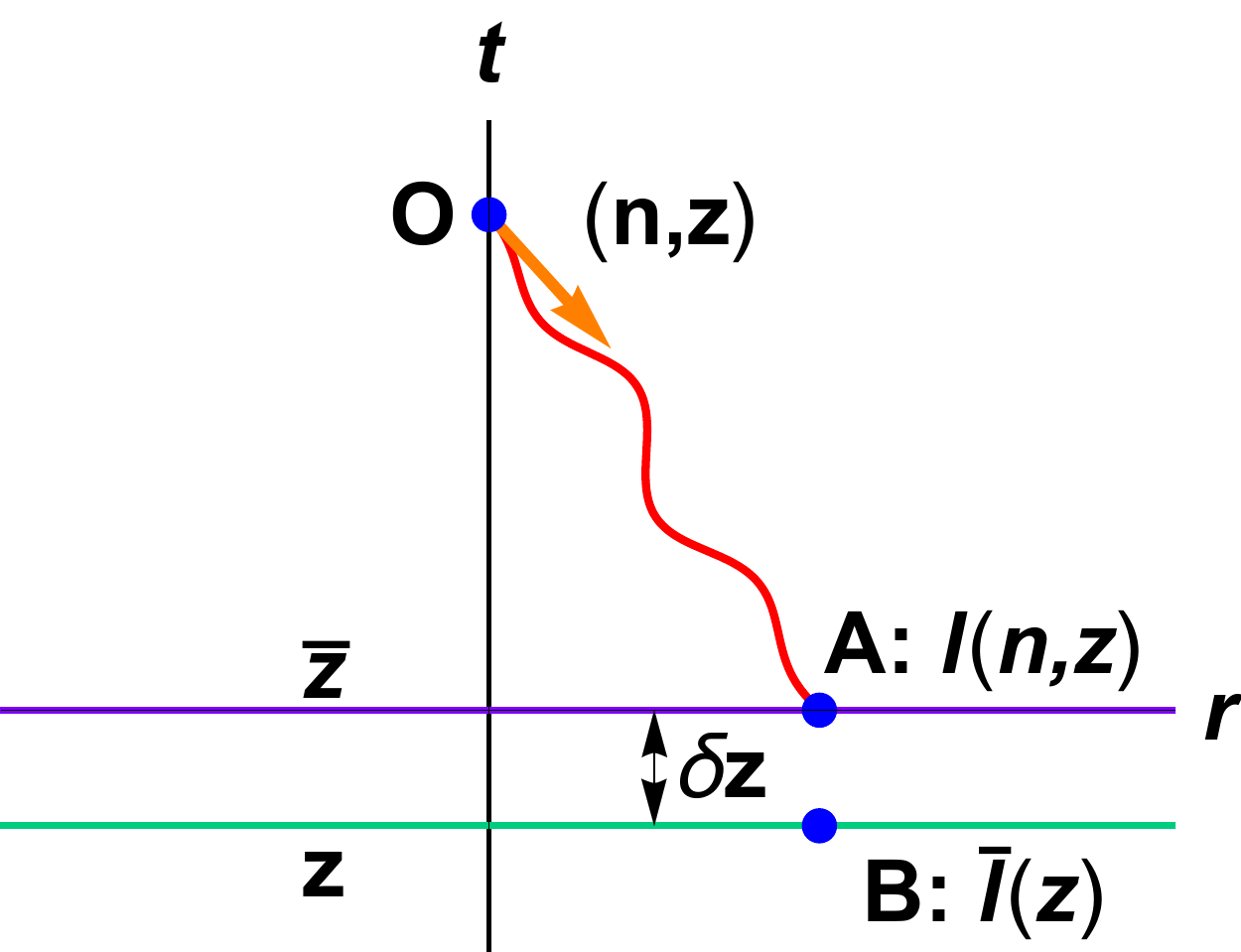}
	\caption{Schematic view of the photon trajectory. The distance is displayed horizontally and the time vertically. The photon is emitted at $A$, on the timeslice corresponding to $z_0$, and is received at $O$.}
	\label{fig:photon_travel}
\end{figure}

The first and second equality are definitions. To go to the second line, we use the \emph{dictionary} to go from Einstein to Jordan frame. To go to the third line, we use that $\ell(\mathbf{n},z) = \ell_{0}(z_0)+\delta \ell(\mathbf{n},z)$ and $z=z_0+\delta z$ to get both contributions. This can be understood by observing figure \ref{fig:photon_travel}. The point $A$ is at the observed coordinate $(\mathbf{n},z)$, but is located on the time slice corresponding to $z_0$. Hence, the perturbation $\delta f$ is given with respect to $f_0(z_0)$.\\
We can understand the two corrections as follows: suppose that $\delta z >0$, as shown in the figure, and that the derivative in the last term is positive, namely $\ell_0$ decreases over time, which is equivalent to saying that the standard ruler is getting larger over time. This implies that length will appear smaller and energies bigger in Einstein frame. Hence the $+n$ in the first term to take into account this correction. Now suppose that $\delta \ell > 0$, which means that lengths appear larger and energy smaller at the considered point that on the time slice, hence the $-n$ of the second term.

The relationship involving the number count in Jordan frame goes as follows:
\begin{align}
\delta_z(\mathbf{n},z) 
&= 
\frac{\rho(\mathbf{n},z)- \rho_0(z)}{\rho_0(z)} \\
&=
\frac{
	\frac{N(\mathbf{n},z) \ell(\mathbf{n},z)^{-1}}{V(\mathbf{n},z)} -
	\frac{N_0(z) \ell_{0}(z)^{-1}}{V_0(z)} 
}
{	\frac{N_0(z) \ell_{0}(z)^{-1}}{V_0(z)}  } \\
&=
\frac{
	\frac{N(\mathbf{n},z)\ell_{0}(z_0)^{-1} \left(1 - \frac{\delta \ell}{\ell_{0}} \right)}{V_0(z) + \delta V (\mathbf{n},z)} -
	\frac{N_0(z) \ell_{0}(z)^{-1}}{V_0(z)} 
}
{	\frac{N_0(z) \ell_{0}(z)^{-1}}{V_0(z)}  } \\
&=
\frac{
	N(\mathbf{n},z)\ell_{0}(z)^{-1}
	\left( 1 + \frac{\delta z}{\ell_{0}} \frac{\mathrm{d}\ell_{0}}{\mathrm{d}z_0} \right) 
	\left(1 - \frac{\delta \ell}{\ell_{0}} \right)\frac{1}{V_0(z)}\left(1 - \frac{\delta V}{V_0}\right) -
	\frac{N_0(z) \ell_{0}(z)^{-1}}{V_0(z)} 
}
{	\frac{N_0(z) \ell_{0}(z)^{-1}}{V_0(z)}  }  \\
&=
\Delta(\mathbf{n},z) +  \frac{\delta z}{\ell_{0}} \frac{\mathrm{d}\ell_{0}}{\mathrm{d}z_0} - \frac{\delta \ell}{\ell_{0}} - \frac{\delta V}{V}.
\label{eq:deltazEApp}
\end{align}
The first equality is the definition of the density perturbation. For the second equality, we use that $\rho = mN/V$ and use the fact that the measured mass of $N$ particles in Einstein frame is proportional to $\ell^{-1}$. Note that we only \emph{convert} the mass, because at the end we would like to obtain the number count, which involves $N$. We do not want to convert the volume here. In the third equality, we use that $\ell(\mathbf{n},z)= \ell(z_0) + \delta \ell(\mathbf{n},z)$ (the logic is the same as in the previous derivation). In the fourth equality, we use $z = z_0 + \delta z$. In the final equality we simplify the expression. 

By using the definition for the redshift density perturbation given in \eqref{eq:deltazEdef}, we can finally write the number counts as

\be
\Delta(\mathbf{n},z) 
=
\frac{\delta \rho}{\bar \rho(z_0)} 
- \frac{\mathrm{d}\rho_0}{\mathrm{d}z_0} \frac{\delta z }{\rho_0(z_0)}
+
\frac{\delta V}{V}
- \frac{\delta z}{\ell_{0}} \frac{\mathrm{d}\ell_{0}}{\mathrm{d}z_0} 
+
\frac{\delta \ell}{\ell_{0}}.
\label{eq:deltazEdefApp}
\ee
Let us present an intuitive explanation for each of these terms. Keep in mind that all the perturbed quantities are evaluated at the point with observed coordinates $(\mathbf{n},z)$, which is physically located at $B$ in figure \ref{fig:photon_travel}. The question we ask is: \emph{Why is the number count not equal to the density perturbation?} Let's consider one term after another. In what follows, we will consider that $\delta z > 0$, as shown for example in the figure \ref{fig:photon_travel}.

\begin{itemize}
    \item The first term is due to the fact the galaxy is located at $z_0$ but we compare it with the background value at the observed redshift, namely $\rho_0( z)$ to obtain the number count. However, $\delta \rho$ is computed from the background value $\rho_0(z_0)$. We have in our example $\rho_0(z_0) < \rho_0(z)$, so the number count will be slightly smaller than the density perturbation. This can be mathematically understood because the derivative of $\rho_0$ in the first term is positive, and we choose $\delta z > 0$ for our example, hence the term is positive but contributes negatively because of the minus sign.
    \item The second term is due to the volume perturbation. Let's suppose that $\delta V>0$, namely the physical volume at the point of the emission is slightly \emph{larger} than on the rest of the time slice. The number count is not affected by any volume distortion, but the density is reduced. Hence, the number count should be bigger than the density perturbation.
    \item The third and the fourth term are the two corrections to the first two terms, their interpretation being given above.
\end{itemize}

\bibliography{refs}
\bibliographystyle{JHEP}
\end{document}